\documentclass[aps,pra,twocolumn,superscriptaddress,groupedaddress,nofootinbib]{revtex4}

\usepackage[pdftex]{graphicx}
\usepackage{graphicx}
\usepackage{dcolumn}%
\usepackage{bm}%
\usepackage{amsmath} 
\usepackage[caption=false]{subfig}
\usepackage{hyperref}
\usepackage{soul}
\usepackage{ulem}

\let\vec\boldsymbol

\begin{document}
\title{Spin polarization of electrons by ultraintense lasers}
\author{D. Del Sorbo}
\affiliation{York Plasma Institute, Department of Physics, University of York, York YO10 5DD, United Kingdom}
\author{D. Seipt}
\affiliation{Cockcroft Institute, Daresbury Laboratory, Warrington WA4 4AD, United Kingdom}
\affiliation{Department of Physics, Lancaster University, Lancaster LA1 4YB, United Kingdom}
\author{T. G. Blackburn}
\affiliation{Department of Physics, Chalmers University of Technology, Gothenburg SE-41296, Sweden}
\author{A. G. R. Thomas}
\affiliation{Department of Physics, Lancaster University, Lancaster LA1 4YB, United Kingdom}
\affiliation{Cockcroft Institute, Daresbury Laboratory, Warrington WA4 4AD, United Kingdom}
\author{C. D. Murphy}
\affiliation{York Plasma Institute, Department of Physics, University of York, York YO10 5DD, United Kingdom}
\author{J. G. Kirk}
\affiliation{Max Planck Institut f\"ur Kernphysik, Saupfercheckweg 1, Heidelberg D-69117, Germany}
\author{C. P. Ridgers}
\affiliation{York Plasma Institute, Department of Physics, University of York, York YO10 5DD, United Kingdom}

\begin{abstract}

Electrons in plasmas produced by next-generation ultraintense lasers ($I> 5\times10^{22}$ W/cm$^{2}$) can be spin polarized to a high degree (10\%-70\%) by the laser pulses on a femtosecond timescale.  This is due to electrons undergoing spin flip transitions as they radiate gamma-ray photons, preferentially spin polarizing in one direction. Spin polarization can modify the radiation reaction force on the electrons, which differs by up to 30\% for opposite spin polarizations. Consequently, the polarization of the radiated gamma-ray photons is also modified: the relative power radiated in the $\sigma$ and $\pi$ components increases and decreases by up to 30\% respectively, potentially reducing the rate of pair production in the plasma by up to 30\%.

\end{abstract}

    \maketitle
  
\section{Introduction}

At the intensities which will be reached by next-generation ultraintense lasers ($\gtrsim 5\times 10^{22}-10^{24}$ W/cm$^{2}$), such as several of those comprising the `Extreme Light Infrastructure' \cite{mourou2007relativistic}, light-matter interactions are predicted to reach the new \color{red} quantum electrodynamic (QED) plasma regime.  \color{black}  Matter in the laser focus is rapidly ionized creating a plasma whose behavior is characterized by the interplay of relativistic plasma and `strong-field' quantum electrodynamic processes \cite{di2012extremely}.  Understanding this interplay is of fundamental interest: this regime is similar to that inferred to exist in extreme astrophysical environments such as pulsar \cite{goldreich1969pulsar} and active black hole magnetospheres \cite{blandford1977electromagnetic}. The QED processes can strongly modify the plasma's behavior, for example leading to complete absorption of the laser pulse, with consequences for potential applications of these lasers, ranging from compact particle accelerators \cite{macchi2013ion} to x-ray source generation \cite{di2012extremely}.   \color{red} Despite its importance, the role of fermion spin in collective high-intensity laser-matter dynamics has rarely been considered \cite{PhysRevE.96.023207}. \color{black}

The important strong-field QED processes in laser-created QED plasmas are \cite{RevModPhys.38.626,di2012extremely,Ritus:JSLR1985}: (i) incoherent emission of $\gtrsim$ MeV energy gamma-ray photons by electrons and positrons on acceleration by the macroscopic electromagnetic fields in the laser-produced plasma (strongly non-linear Compton scattering), with the resulting radiation-reaction strongly modifying the dynamics of the emitting electron or positron \cite{ridgers2014modelling,di2010quantum}; (ii) pair creation by the emitted gamma-ray photons, in the same electromagnetic fields (the multi-photon Breit-Wheeler process \cite{burke1997positron}). For example: non-linear Compton scattering and the resulting radiation can lead to almost complete laser absorption \cite{zhang2015effect,seipt2016depletion}; pair cascades (where pairs emit further  gamma-ray photons, which generate even more pairs), can lead to the creation of critical density pair plasmas \cite{grismayer2016laser,Nerush:PRL2011,ji2014energy,del2017efficient}.  It is therefore essential that we correctly include these QED processes in our models of the interaction of next-generation laser pulses with matter.
Previous treatments of gamma-ray photon emission in QED plasmas have averaged over the spin. 

 In this \color{red} article \color{black} we demonstrate a novel process, analogous to the Sokolov-Ternov effect in a strong magnetic field \cite{sokolov1966synchrotron,belomesthnykh1984observation}, whereby the electrons in laser-generated QED plasmas rapidly spin polarize due to asymmetry in the rate of spin flip transitions, i.e. interactions where the spin changes sign during the emission of a gamma-ray photon.  We discuss several signatures of the spin polarization of the plasma. (i) Consideration of the spin of the electrons leads to a new quantum correction to the radiation reaction force: in the energetically favourable spin configuration the total power radiated and so radiation reaction force is enhanced compared to the unfavourable configuration, leading to an enhancement in total power radiated by the plasma compared the prediction assuming the spin is unpolarized.  
(ii) The relative energy emitted in the two possible polarizations of the gamma-ray photons is modified.  (iii) This modification could decrease the rate of pair production.  


\color{red}

Spin polarized electron beams are important in high energy physics; the use of properly polarized beams in electron-positron colliders can suppress the Standard Model background in searches for new physics beyond the Standard Model \cite{vauth2016beam}. Spin polarized beams are also used in electron spectroscopy for studying surface \& thin-film magnetism and the electronic structure of metal, semiconductor surfaces and films \cite{subashiev1998spin,siegmann1992surface}. 
Spin polarizing plasma may enable applications of ultraintense lasers in these areas.

The paper is organized as follows. In Sec.~\ref{Spin polarization by laser pulses} we demonstrate the laser-induced process of electron spin polarization by deriving a simple predictive model and applying it to the case of electrons orbiting at the magnetic node in the field of two counter-propagating lasers. In Sec.~\ref{Consequences of spin polarization} we discuss signatures \& consequences of spin polarization. Finally, in Sec.~\ref{Conclusions} we draw conclusions.\color{black}

\section{Spin polarization by laser pulses}\label{Spin polarization by laser pulses}

We focus on the case of electrons orbiting in a rotating electric field -- a configuration that may be realized in the plasma created at the magnetic node of two colliding, circularly-polarized laser pulses.  The electric field $\vec{E}$ at the magnetic node, say the plane $z=0$, rotates with constant amplitude \cite{bell2008possibility}. Consequently, electrons subjected to $\vec{E}$ also rotate in the plane
$z=0$ with (normalized) velocity $\vec{\beta}$.

To describe the spin polarization dynamics of electrons in the rotating electric field (indeed in any arbitrary electromagnetic field)
we need to use a proper non-precessing spin polarization basis $\vec \zeta$
for which $d \vec \zeta/d\tau = 0$ \cite{king2015double}, defined in the rest frame of the electron, where $\tau=t/\gamma$ is the proper time.  Spin-up and spin-down
electron states, defined with respect to that basis, do not mix.
According to the Bargmann-Michel-Telegdi equation \cite{bargmann1959precession}, which describes the precession of the expectation value of the spin polarization vector in an external electromagnetic field \cite{vieira2011polarized}, the only non-precessing spin basis for an electron rotating in the rotating electric field is 
\begin{equation}
\vec\zeta = \frac{\vec{E}\times\vec{\beta}}{\Vert\vec{E}\times\vec{\beta}\Vert} = \vec e_z.
\end{equation}
 We will therefore consider spin polarization in the transverse direction where $\vec \zeta$
is perpendicular to $\vec{\beta}$. $\vec{\zeta}$ is analogous to the direction along the magnetic field in the discussion of the Sokolov-Ternov effect. 
%


%

Each electron in the rotating electric field has a projection of its spin in the direction of the vector $\vec \zeta$ of $s \hbar /2$, where $s=\pm1$ and $\hbar$ is Planck's constant. We therefore divide the plasma electron
population into two fractions $n^s$, characterized by their spin projection being
parallel ($s=1=\uparrow$, higher energy) or antiparallel ($s=-1=\downarrow$, lower energy)
to $\vec\zeta$.
The number density of electrons with spin $s=\downarrow$, for instance, evolves according to the master equation
\begin{align} \label{eq:master}
\frac{d}{d\tau}n^{\downarrow}(t) = \frac{dN^{\uparrow\downarrow}}{d\tau}n^{\uparrow}(t)-\frac{dN^{\downarrow\uparrow}}{d\tau}n^{\downarrow}(t).
\end{align}
The  gamma-ray photon emission rates
${dN^{ss'}}/{d\tau} = \sum_\rho {dN^{ss'}_\rho}/{d\tau} $,
summed over photon polarization states $\rho$, describe the transitions from one spin state to the other on emission of a gamma-ray photon, i.e. spin flip transitions. 
An analogous equation holds for $n^{\uparrow}$, the number density of electrons with spin $s=\uparrow$.

\begin{figure*}
\subfloat[\label{W.pdf}]{%
        \centering
        \includegraphics[width=0.85\columnwidth]{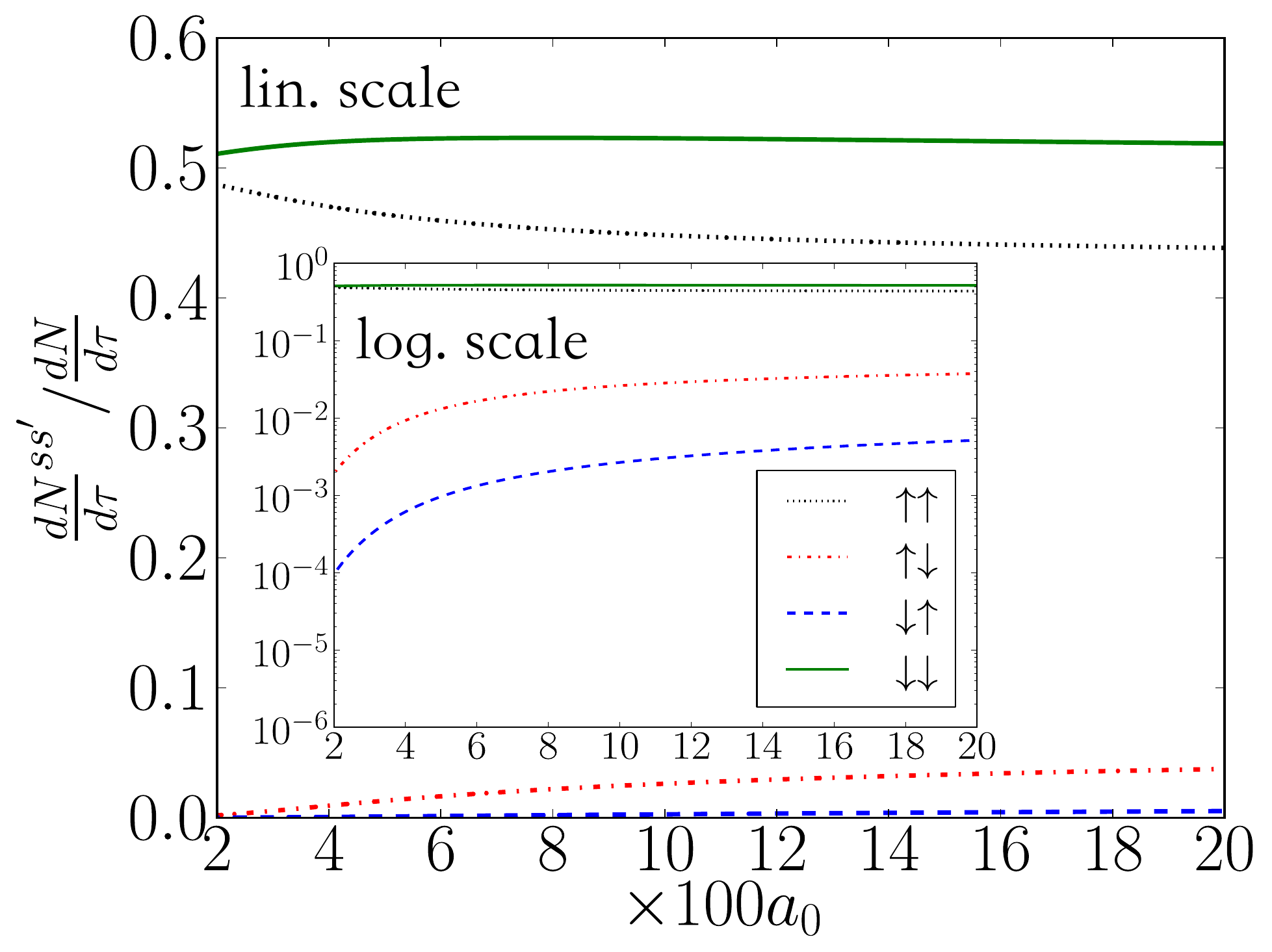}}
~
\subfloat[\label{Magnetization.pdf}]{%
        \centering
        \includegraphics[width=0.9\columnwidth]{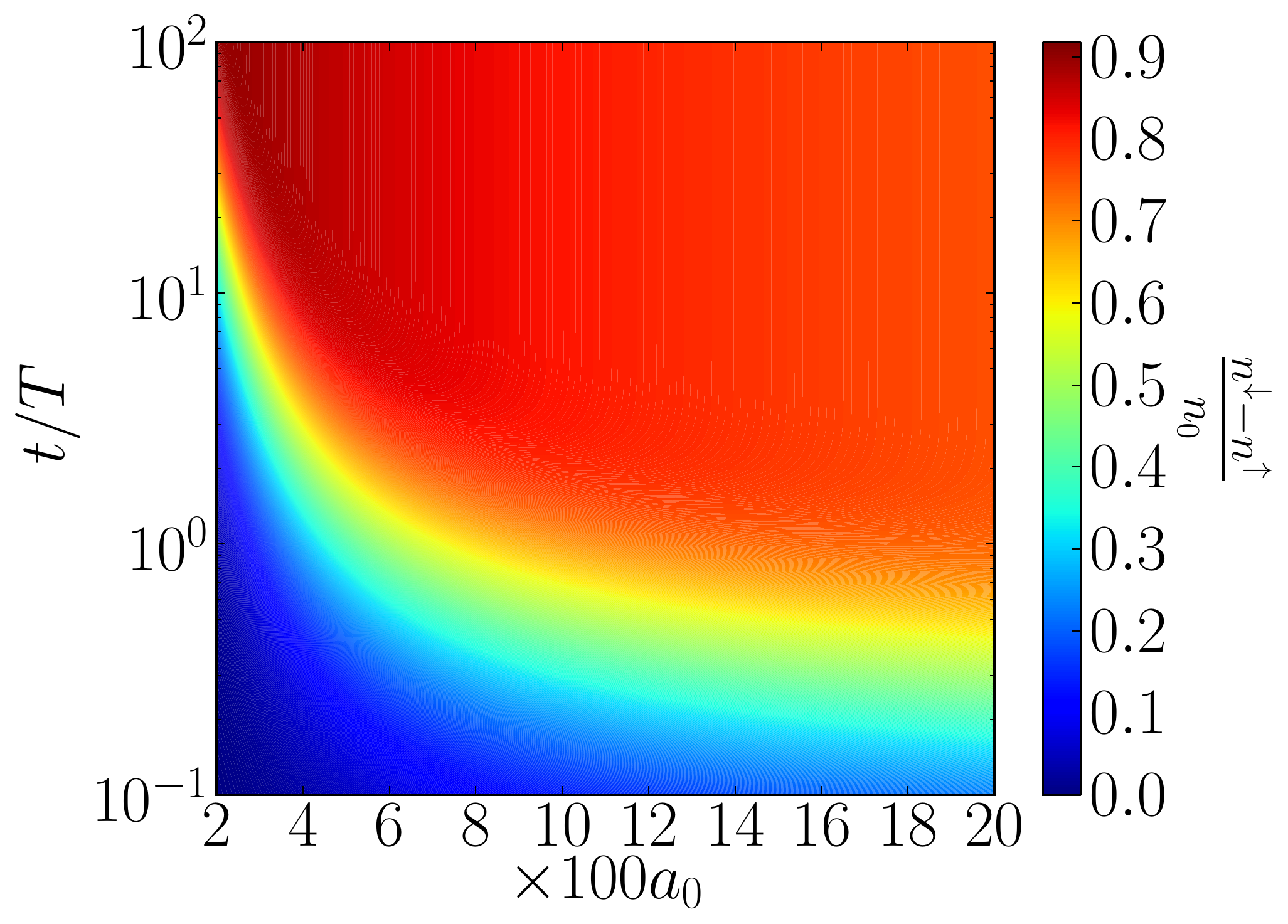}
                        }
      \caption{(a) The rates in Eq.~\eqref{transition probability B}, summed over photon polarization and normalized to the unpolarized rate ${dN}/{d\tau}$, as functions of the strength parameter of the laser electromagnetic waves $a_{0}$. 
 (b) Degree of electron spin polarization antiparallel, as a function of $a_{0}$ and of time normalized to the laser period $T\approx3.33$ fs. 
 }
\end{figure*}

\subsection{Rate of spin flip transitions}

The expressions for the rate of spin flip transitions can be simplified when the strength parameter $a_0 = 85.5 \lambda/\mu \rm m\sqrt{I/(10^{22} \; W/cm^{2})}$ of the laser electromagnetic waves is large. In that case, the formation phase interval for the emission of a photon becomes very short, $\propto 1/a_0$, and multiple photon emissions can be considered as incoherent. $a_0 $ is indeed large in the cases considered here, where the intensity of each of the counter-propagating laser pulses $I$ is  $\gg 10^{18}$  W/cm$^{2}$ and $\lambda = 1 \;\mu$m -- typical wavelength of ultraintense lasers and assumed in all our calculations. It is also the case that the electric field at the magnetic node is much less than the critical field of QED: the Sauter-Schwinger field $E_{S}=m_{e}^{2}c^{3}/(e\hbar)$ \cite{heisenberg1936folgerungen}, where $m_e$ and $e$ are the electron mass and charge respectively, and $c$ is the speed of light (all in Gaussian units). Under these two assumptions t\color{black}he rates of gamma-ray photon emission (and pair production) are well described by the corresponding rates in
constant crossed electric and magnetic fields \cite{Ritus:JSLR1985,sokolov1966synchrotron,baier1968processes,kirk2009pair}. 
The rates of photon emission in this constant crossed field approximation depend only on: the electron's initial and final spin projections $s$ and $s'$, the emitted photon's polarization $\rho$ and the quantum efficiency parameter $\eta=E_{RF}/E_{S}$ \cite{baier1968processes}.
Here,
$E_{RF}=\gamma \Vert \vec{E} \Vert$
is the electric field's magnitude in the electron's instantaneous rest frame.  The electron spin and photon polarization dependent photon emission rates read  \cite{sokolov1966synchrotron,bordovitsyn1999synchrotron}
\begin{equation}
\frac{dN^{ss'}_{\rho}}{d\tau} ={\mathcal{P}_\mathrm{class}}\int_{0}^{\eta/2}d\chi\frac{dy}{d\chi}\frac{F(\eta,\chi,s,s',\rho)}{\hbar\omega/\gamma},\label{transition probability B}
\end{equation}
with the classical dipole radiation emission power 
\begin{equation}
\mathcal{P}_\mathrm{class}=\frac{2m_{e}^{2}c^{3}e^{2}\eta^{2}}{3\hbar^{2}}.
\end{equation}
The photon quantum parameter 
\begin{equation}
\chi=\frac{\hbar\omega E}{2m_{e}c^{2}E_{S}}
\end{equation}
 depends on the gamma-ray photon energy 
 \begin{equation}
 \hbar\omega=\frac{\gamma m_{e}c^{2}\xi y}{1+\xi y}, 
 \end{equation}
 with $\xi={3\eta}/{2}$ and 
 \begin{equation}
 y=\frac{4\chi}{3\eta(\eta-2\chi)}.
 \end{equation} 
$F(\eta,\chi,s,s',\rho)$ is the spin- and photon polarization-dependent quantum synchrotron function. It is given by: 
\begin{widetext}
\begin{multline}
F(\eta,\chi,s,s',\rho)=\frac{9\sqrt{3}y}{16\pi(1+\xi y)^{4}}\Bigg\{\frac{1+ss'}{2}
\bigg\{ \left(1+\frac{1}{2}\xi y\right)^{2}\left[\int_{y}^{\infty}K_{5/3}(x)dx+  \rho  K_{2/3}(y)\right]\\+  \left[\frac{\xi^{2} y^{2}}{2}\int_{y}^{\infty}K_{1/3}(x)dx-s'(2+\xi y)\xi yK_{1/3}(y) \right]\frac{1+ \rho }{2}\bigg\}\\+  \frac{1-ss'}{2}\frac{\xi^{2} y^{2}}{4}\bigg\{\int_{y}^{\infty}K_{5/3}(x)dx-\rho K_{2/3}(y)+\left[2\int_{y}^{\infty}K_{1/3}(x)dx-4s'K_{1/3}(y)\right]\frac{1-\rho}{2}\bigg\}
\Bigg\},
\end{multline}
\end{widetext}
 where $K$ as the modified Bessel function of the second kind.  The variable $\rho$ refers to the photon polarization, where $\rho=1=\sigma$ (polarized orthogonal to $\vec\zeta$) and $\rho=-1=\pi$ (polarized in the direction of $\vec\zeta$).

Fig.~\ref{W.pdf} shows the rates in Eq.~\eqref{transition probability B}, summed over photon polarizations,  $dN^{ss'}/d\tau= dN^{ss'}_{\sigma}/d\tau+dN^{ss'}_{\pi}/d\tau$, 
as functions of laser power of one of the counter-propagating pulses (the rates depends on $\eta$ which in turn depends on $a_{0}$, as explained in the next paragraph) and normalized to the unpolarized rate  for a single electron  ${dN}/{d\tau}= \sum_{ss'}dN^{ss'}/d\tau$ \footnote{Our ${dN}/{d\tau}$ is twice that defined in Refs.~\cite{kirk2009pair,ridgers2014modelling}, because two electron populations, characterized by their polarizaton, are involved. }.
The rate of processes resulting in a final spin aligned antiparallel to
$\vec\zeta$
(${dN^{\downarrow\downarrow}}/{d\tau}+{dN^{\uparrow\downarrow}}/{d\tau}$) is higher than those resulting in a final spin aligned parallel 
(${dN^{\uparrow\uparrow}}/{d\tau}+{dN^{\downarrow\uparrow}}/{d\tau}$) due to the difference in the rates of spin flip. This means that the electron spin tends to align itself antiparallel to 
$\vec\zeta$ on a time scale which we denote as the spin polarization time $t_\mathrm{pol}$.

\begin{figure*}
\subfloat[\label{Intensity.pdf}]{%
        \centering
        \includegraphics[width=0.9\columnwidth]{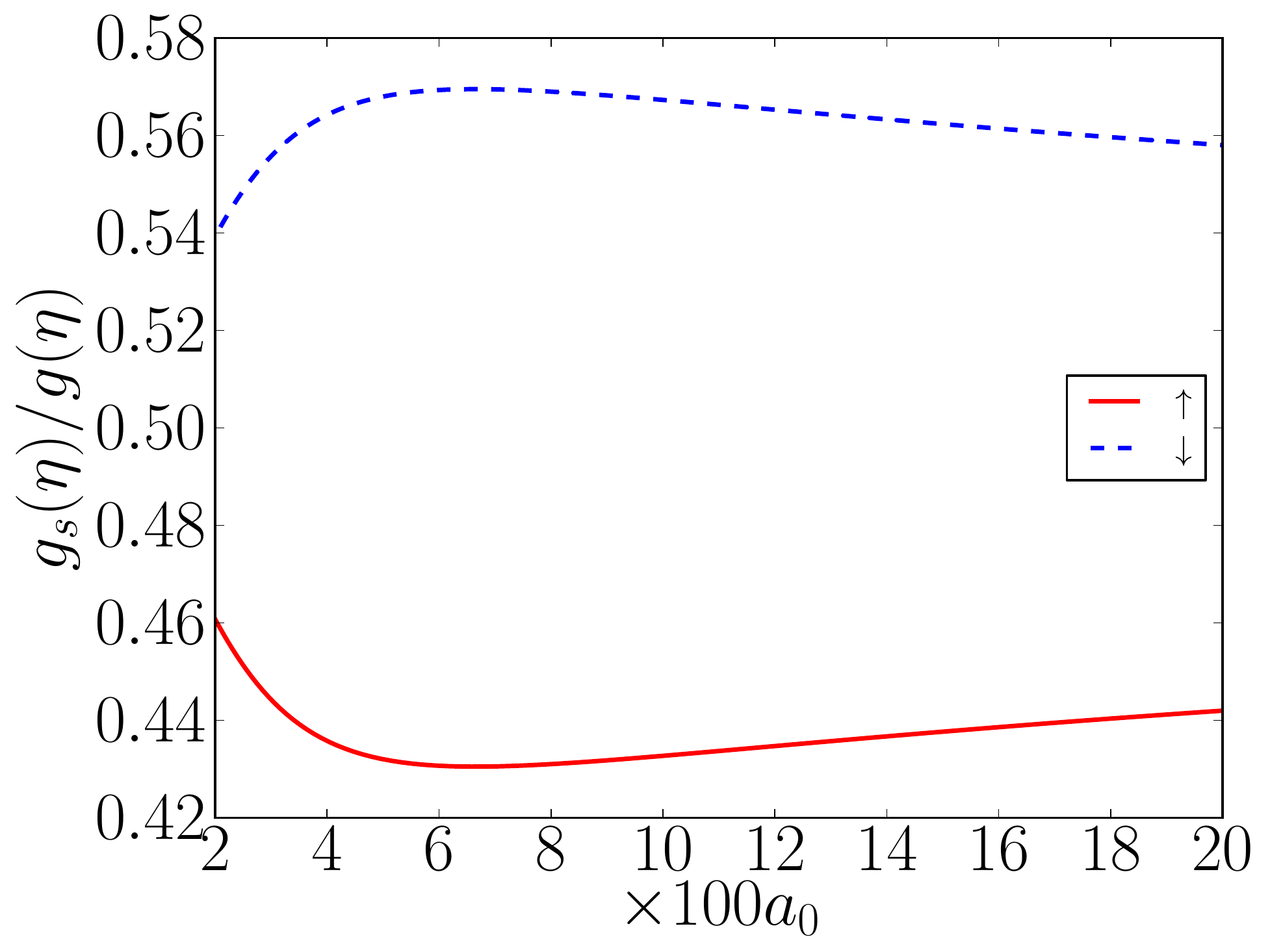}}
~
\subfloat[\label{Photon_pol.pdf}]{%
        \centering
        \includegraphics[width=0.9\columnwidth]{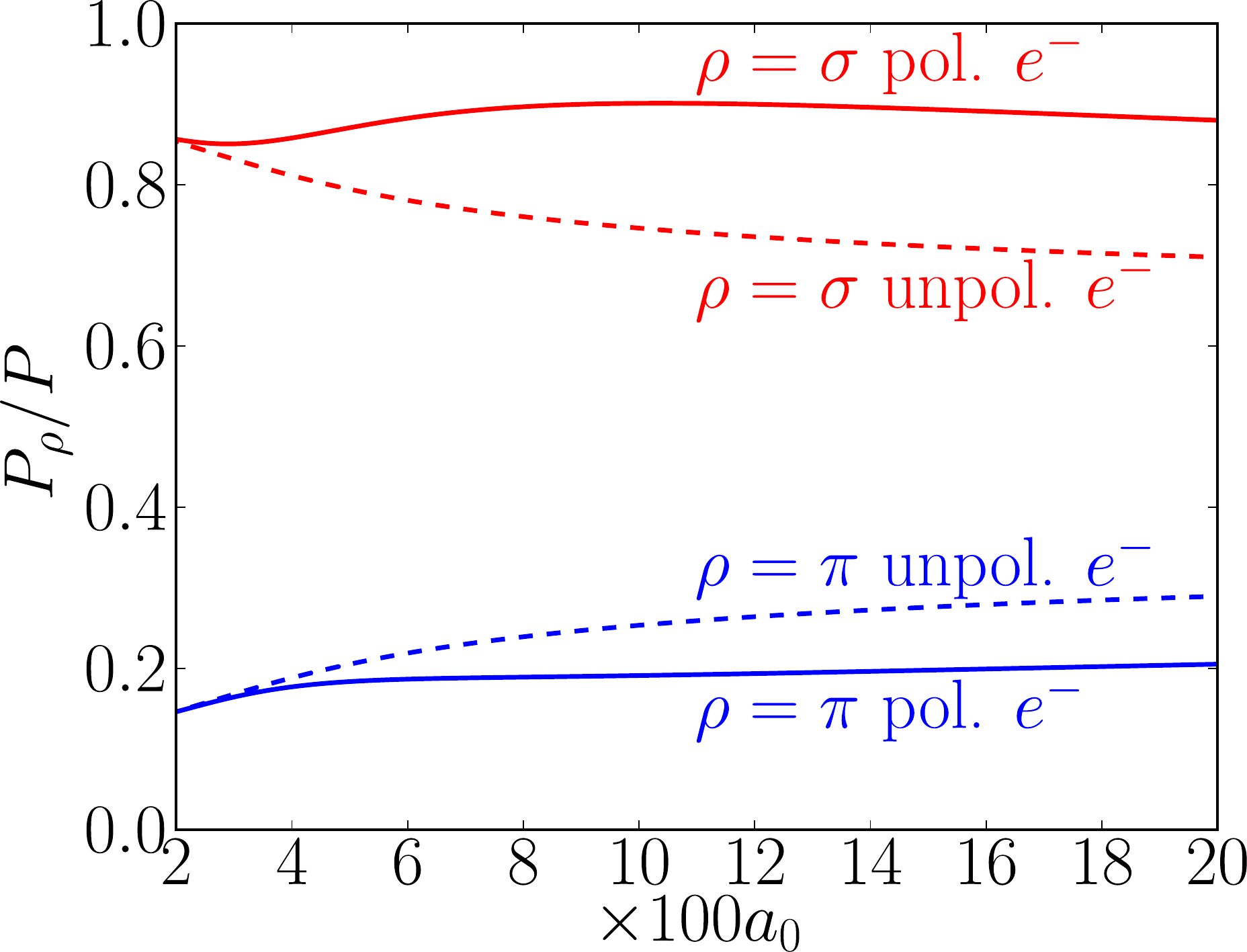}
                        }
      \caption{(a) Spin-dependent $g^{s}(\eta)$ factor normalized to its unpolarized analogous. (b) Power radiated by polarized electrons in both $\sigma$ and $\pi$ polarizations,
  at $t=T$, normalized to the total power emitted by unpolarized electrons.
      All the quantities are plotted as functions of $a_{0}$.}
\end{figure*}

In order to calculate the rates in figure \ref{W.pdf} we had to determine $\eta$.  The electrons rotating in the electric field reach a steady state when the radiative losses due to gamma photon emission balance the acceleration due to the electric field.  In this case \cite{zhang2015effect} $\eta\approx 206\varepsilon_\mathrm{rad} \gamma^{2}$,
where 
\begin{equation}
\varepsilon_\mathrm{rad}= { \frac{4\pi e^2}{3m_e c^2\lambda} } \approx 1.18 \times 10^{-8}
\end{equation}
 and the electron's Lorentz factor is given by 
\begin{equation}
\left[g(\eta) \, \varepsilon_\mathrm{rad} \gamma^{4}\right]^{2}+\gamma^{2}= a_{0}^{2}.
\end{equation}
The factor
\begin{equation}
g(\eta)=\sum_{ss'\rho}\frac{\mathcal{P}^{ss'}_{\rho}}{\mathcal{P}_\mathrm{class}}=\sum_{ss'\rho}\int_{0}^{\eta/2}d\chi\frac{dy}{d\chi}F(\eta,\chi,s,s',\rho)
\end{equation}
 accounts for the reduction of the power radiated due to quantum effects \cite{ridgers2014modelling},
as compared to classical emission, where $\mathcal{P}^{ss'}$ is the power radiated during an emission where the initial and final spins are $s$ and $s'$ respectively.



The Sokolov-Ternov effect in a magnetic field has been been observed after one hour in storage rings \cite{sokolov1966synchrotron}, for which  the  quantum efficiency parameter $\eta\sim10^{-6}$ or less \cite{mane2005spin}. Next-generation ultraintense lasers  are expected to reach more extreme regimes, in which $\eta\sim0.1-1$ \cite{bell2008possibility}. These high values of $\eta$ can be also achieved in the interaction of relativistic particles with strong crystalline fields \cite{uggerhoj2005interaction}, as it has been experimentally observed \cite{andersen2012experimental}.

\subsection{Spin polarization time}

We may determine the degree of spin polarization antiparallel to $\mathbf{E\times\beta}$ (i.e. $\downarrow$) in time $t$ by solving \eqref{eq:master} with the rates \eqref{transition probability B}.  We find that it evolves as
\begin{equation}
\frac{n^{\downarrow}-n^{\uparrow}}{n_{0}}
= \frac{\frac{dN^{\uparrow\downarrow}}{dt} -\frac{dN^{\downarrow\uparrow}}{dt}}
			{\frac{dN^{\uparrow\downarrow}}{dt}+\frac{dN^{\downarrow\uparrow}}{dt}}
			\left(1-e^{-t/ t_\mathrm{pol}}\right) \,,\label{polarizzazione}
\end{equation}
with the spin polarization time
$t_\mathrm{pol}=\gamma({dN^{\uparrow\downarrow}}/{d\tau}+{dN^{\downarrow\uparrow}}/{d\tau})^{-1}$,
and assuming initially unpolarized electrons with total density $n_0 = n^\uparrow + n^\downarrow$. 
By inverting the signs of $s$ and $s'$ this equation describes the polarization of positrons as well \cite{berestetskii1982quantum}.

In Figure \ref{Magnetization.pdf} we have plotted the solutions of Eq.~\eqref{polarizzazione} as a function of time, normalized to the laser period $T=\lambda/c\approx3.33$ fs, and of $a_{0}$. We see that for laser intensities just beyond the current limit ($a_{0}\approx200$; $I\approx5\times10^{22}$ W/cm$^{2}$) a significant ($\approx10\%$) spin polarization is expected to occur rapidly, i.e. within a single laser period. $\gtrsim50\%$ spin polarization is expected in one laser period for lasers of intensity well within the reach of next-generation laser systems
($a_{0}\gtrsim600$; $I\gtrsim5\times10^{23}$ W/cm$^{2}$) and a maximum spin
polarization of 70\% after one laser period is expected at a laser intensity $5\times10^{24}$ W/cm$^{2}$ ($a_{0} \approx 2000$).

The orbit considered here is unstable on a timescale of the order of the laser period \cite{kirk2016radiative}. However, as we have shown, a high degree of spin polarization can indeed occur in such a short timescale. 

\section{Consequences of spin polarization}\label{Consequences of spin polarization}

Let us now discuss several immediate consequences of the electron polarization on the subsequent QED plasma dynamics.
Figure \ref{Intensity.pdf} shows the  spin-dependent $g^{s}(\eta)=\sum_{s'\rho}\mathcal{P}^{ss'}_{\rho}/\mathcal{P}_\mathrm{class}$ by polarized electrons.  The intensity of emission from an electron initially in the state $s=\downarrow$ is up to 30\% higher than that from an electron initially in the state $s=\uparrow$. Hence, the power radiated and, consequently, the radiation reaction force depend on the spin of the electron.  This is currently not included in the modeling of high intensity laser-matter interactions.  As a result complete spin polarization of the plasma will increase the total power radiated by the plasma by up to $15\%$ compared to the emission from an equivalent unpolarized plasma; this could provide an observable signature of the spin polarization.
Moreover, we would expect this 
causes an increased rate of laser absorption.

 Note that the effect of spin on $\eta$ and $\gamma$ through the spin-dependent $g(\eta)$ is smaller than on the total power radiated ($<5\%$ instead of 15\%, as can be numerically tested). Therefore, it was neglected in the calculation of the degree of spin polarization.  


 The electron spin polarization also affects polarization of the emitted gamma-ray photons.
The power radiated in both $\sigma$ and $\pi$ polarizations,
$P_{\sigma,\pi}=\sum_{ss'}n^{s}\mathcal{P}^{ss'}_{\sigma,\pi}$,
  at $t=T$, is plotted in Fig.~\ref{Photon_pol.pdf}, normalized to the total power emitted by unpolarized electrons. Electron spin polarization causes the relative power radiated as $\sigma$ photons to increase by up to 30\% compared to the unpolarized case (vice versa for $\pi$ photons). The relative yields of $\sigma$ and $\pi$ photons provides another signature of the spin polarization of the plasma.

The gamma-ray photons emitted by the electrons can decay to pairs in the electromagnetic fields of the laser pulses by the multi-photon Breit-Wheeler process. 
 Pair cascades become important rapidly as $a_{0}$ exceeds 1200 ($I=2\times10^{24}$ W/cm$^{2}$) \cite{bell2008possibility,fedotov2010limitations,Bulanov:PRA2013,grismayer2015seeded}. 
 The polarization of the gamma-ray photons has been shown to modify the rate of pairs produced by up to 30\% \cite{king2013photon}.  The modification to the polarization of the gamma-ray photons, caused by the spin polarization of the electrons,
 would be expected to produce a reduction to the rate of pair production of the same order and thus both electron and positron spin and photon polarization must be included in cascade simulations, yet they are currently neglected.
The generated positrons will also spin polarize (parallel to $\vec\zeta$). 
Although prolific pair production by a cascade will add source terms to equations for $dn^{\uparrow}/d\tau$ \& $dn^{\downarrow}/d\tau$ (plus two additional equations for positrons), we would not expect a qualitative change to the spin polarization trend in the magnetic node.   Indeed, the rapid increase in the rate of pair production with laser intensity means that either the interaction is in a regime where a cascade does not occur and the number of positrons is small, or the cascade is rapidly quenched as the number density of pairs reaches the relativistic critical density and the source term shuts off.

The spin magnetization of electrons in a plasma at the magnetic node, resulting from the spin polarization,  can be deduced by multiplying Eq.~\eqref{polarizzazione} by $\mu_{B}n_{0}$ \cite{jackson1999classical}, where $\mu_{B}$ is the Bohr magneton.
Assuming the plasma density is equal to the relativistic critical density (upper limit for laser propagation), the spin magnetization for $1$ $\mu$m wavelength lasers with $I>5\times10^{22}$ W/cm$^{2}$ ($a_{0}>200$) is $M\sim\rm kG$, after one laser period. This quantity may be considered another observable effect of spin polarization.

\section{Conclusions}\label{Conclusions}
We have shown that electrons in a plasma created by two counter-propagating, ultraintense ($a_{0}>200$; \color{red} $I>5\times10^{22}$ W/cm$^{2}$) \color{black} laser pulses can spin polarize (to $10\%-70\%$) on a femtosecond timescale.
\color{red} In this laser configuration, assuming the plasma is sufficiently rarefied that collective effects are negligible, the only experimental parameter that influence the degree of spin polarization is the laser intensity, since the wavelength is currently fixed to $\sim 1 \;\mu$m by technological limitations. \color{black}

 \color{red} Spin polarization \color{black} can enhance the radiation reaction force on electrons and positrons in the plasma by up to 15\%, compared to the prediction for unpolarized electrons.  Consequently, the power radiated by the plasma is enhanced by the same percentage.
The polarization of the radiated gamma-ray photons is also modified, by 30\%,
  potentially reducing the number of pairs produced in the plasma by 30\%.  
  
  The spin polarization must therefore be accounted for in the modeling of next-generation laser matter interactions.  The possibility of producing spin polarized electrons and positrons with ultraintense lasers also opens up new applications. Polarized electrons are fundamental for the study of particle physics and are used in the spin polarized electron spectroscopy. 

\section*{Acknowledgements}  

This work was funded by the UK Engineering and Physical Sciences Research Council (EP/M018156/1), by the Science and Technology Facilities Council (ST/G008248/1)  \color{red}  and by the Knut and Alice Wallenberg Foundation. The authors are grateful to Antonino Di Piazza for useful and stimulating discussions. The data required to reproduce the results in this paper is available at DOI 10.15124/1afd25a0-a1e3-49ec-afa8-2e5f6d868124. \color{black}


 \end{document}